\documentclass[apj]{emulateapj}
\usepackage{rotating}
\usepackage{amsmath} 
\usepackage{amssymb}
\usepackage{xcolor}
\usepackage{hyperref}
\usepackage{tabularx}
\usepackage{graphicx} 
\usepackage{threeparttable}  
\usepackage{xfrac} 
\usepackage{CJKutf8} 
\usepackage{lineno}

\def\ltsima{$\; \buildrel < \over \sim \;$}
\def\simlt{\lower.5ex\hbox{\ltsima}}
\def\gtsima{$\; \buildrel > \over \sim \;$}
\def\simgt{\lower.5ex\hbox{\gtsima}}
\def\kpc{{\rm\,kpc}}

\def\lsun{{\rm\,L_\odot}}

\def\pc{{\rm\,pc}}

\def\deg{^\circ}

\def\s{\ifmmode \widetilde \else \~\fi}
\def\={\overline}

\def\spose#1{\hbox to 0pt{#1\hss}}

\def\eg{{e.g.,\ }}
\def\ie{{ i.e.,\ }}
\def\lta{\mathrel{\spose{\lower 3pt\hbox{$\mathchar"218$}}
     \raise 2.0pt\hbox{$\mathchar"13C$}}}
\def\gta{\mathrel{\spose{\lower 3pt\hbox{$\mathchar"218$}}
     \raise 2.0pt\hbox{$\mathchar"13E$}}}
\def\Dt{\spose{\raise 1.5ex\hbox{\hskip3pt$\mathchar"201$}}}    
\def\dt{\spose{\raise 1.0ex\hbox{\hskip2pt$\mathchar"201$}}}    

\def\dotsfill{\leaders\hbox to 1em{\hss.\hss}\hfill}

\shorttitle{Global properties of Andromeda's satellite system}
\shortauthors{A. Doliva-Dolinsky et al.}

\begin{document}


\title{The PAndAS View of the Andromeda Satellite System. IV Global properties}


\author{Amandine Doliva-Dolinsky$^{1}$, Nicolas F. Martin$^{1,2}$, Zhen Yuan$^{1}$ (\begin{CJK}{UTF8}{gbsn}袁珍),\end{CJK}Alessandro Savino$^{3}$, Daniel R. Weisz$^{3}$, Annette M. N. Ferguson$^{4}$, Rodrigo A. Ibata$^{1}$, Stacy Y. Kim$^{5}$, Geraint F. Lewis$^{6}$, Alan W. McConnachie$^{7}$, Guillaume F. Thomas$^{8,9}$ }
\email{amandine.doliva-dolinsky@astro.unistra.fr}

\altaffiltext{1}{Universit\'e de Strasbourg, CNRS, Observatoire astronomique de Strasbourg, UMR 7550, F-67000, France}
\altaffiltext{2}{Max-Planck-Institut f\"ur Astronomie, K\"onigstuhl 17, D-69117, Heidelberg, Germany}
\altaffiltext{3}{Department of Astronomy, University of California, Berkeley, CA 94720, USA}
\altaffiltext{4}{Institute for Astronomy, University of Edinburgh Royal Observatory, Blackford Hill, Edinburgh EH9 3HJ, UK}
\altaffiltext{5}{Department of Physics, University of Surrey, Guildford, GU2 7XH, United Kingdom}
\altaffiltext{6}{Sydney Institute for Astronomy, School of Physics, A28, The University of Sydney, NSW 2006, Australia}
\altaffiltext{7}{NRC Herzberg Astronomy and Astrophysics, 5071 West Saanich Road, Victoria, BC, V9E 2E7, Canada}
\altaffiltext{8}{Instituto de Astrof\'isica de Canarias, Calle Vía L\'actea, s/n, 38205. La Laguna, Santa Cruz de Tenerife, Spain}
\altaffiltext{9}{Departamento de Astrof\'isica, Universidad de La Laguna, 38206, La Laguna, Tenerife, Spain}

\begin{abstract}
We build a statistical framework to infer the global properties of the satellite system of the Andromeda galaxy (M31) from the properties of individual dwarf galaxies located in the Pan-Andromeda Archaelogical Survey (PAndAS) and the previously determined completeness of the survey. Using forward modeling, we infer the slope of the luminosity function of the satellite system, the slope of its spatial density distribution, and the size-luminosity relation followed by the dwarf galaxies. We find that the slope of the luminosity function is $\beta=-1.5\pm0.1$. Combined with the spatial density profile, it implies that, when accounting for survey incompleteness, M31 hosts $92_{-26}^{+19}$ dwarf galaxies with $M_\textrm{V}<-5.5$ and a sky-projected distance from M31 between 30 and 300\kpc. We conclude that many faint or distant dwarf galaxies remain to be discovered around Andromeda, especially outside the PAndAS footprint. Finally, we use our model to test if the higher number of satellites situated in the hemisphere facing the Milky Way could be explained simply by the detection limits of dwarf galaxy searches. We rule this out at $>99.9\%$ confidence and conclude that this anisotropy is an intrinsic feature of the M31 satellite system. The statistical framework we present here is a powerful tool to robustly constrain the properties of a satellite system and compare those across hosts, especially considering the upcoming start of the Euclid or Rubin large photometric surveys that are expected to uncover a large number of dwarf galaxies in the Local Volume.

\end{abstract}

\keywords{Local Group --- dwarf galaxies --- Andromeda}

\section{Introduction}
During the last decades, faint dwarf galaxies ($L<10^6\lsun$) have proven to be powerful testbeds for cosmological and galaxy formation models. The majority of these constraints are obtained from the dwarf galaxy satellite system of the Milky Way (MW) \citep[\eg][]{Koposov2009,Kim2018,Nadler2021} because of the difficulty to detect those faint objects beyond our immediate surroundings with current panoptic photometric surveys \citep[\eg][]{Koposov2008,Drlica2020_1}. However, as the Milky Way satellites and past satellite accretion may not be typical \citep[\eg][]{Martin2017,Weisz2019,Evans2020}, it is important to explore the satellite systems of other similar hosts, the most accessible of which is the Andromeda galaxy (M31).  

M31 and the cohort of dwarf galaxies that inhabit its halo are close enough \citep[$\sim800\kpc$;][]{Savino2022} to be resolvable into stars with modern observing capabilities. At the turn of the century, systematic efforts were undertaken to survey the surroundings of our cosmic neighbor, with the Sloan Digital Sky Survey \citep[SDSS][]{SDSS2003} and, more particularly, with the Pan-Andromeda Archaeological Survey \citep[PAndAS][]{McConnachie2018}, the sample of known dwarf galaxies near M31 has increased significantly. From about 10 dwarf galaxies known at the end of the twentieth century \citep{Hershel1789,Vandenbergh1972,Karachentsev1999}, we now know of $\sim40$ dwarf galaxies that are likely satellites of Andromeda. Four of those were discovered from SDSS photometry \citep{Zucker2004,Zucker2007,Bell2011,Slater2011}, four from more localized efforts \citep{Majewski2004,Irwin2008,Collins2022,Martinez-Delgado2022}, three from searches based on the Pan-STARRS1 survey \citep{Martin2013a,Martin2013b}, and, mainly, 19 from the exploration of the deeper PAndAS data \citep{Martin2006,Ibata2007,McConnachie2008,Martin2009,Richardson2011}. Because they are significantly more distant than their MW counterparts, these newly discovered dwarf galaxies are also somewhat brighter than the faintest MW dwarf galaxy satellites but they nevertheless reach total luminosities as faint as $10^{4.2\pm0.4}\lsun$ ($M_V=-6.0^{+0.7}_{-0.5}$ for And~XXVI; \citealt{Savino2022}).

Beyond the mere discovery of satellites, it is essential to also quantify the completeness of those large surveys \citep{Koposov2008,Drlica2020_1} as these are key ingredients to properly fold in observational biases when comparing the known dwarf galaxies (or dwarf galaxy systems) between themselves or with simulations of the faint-end of galaxy formation in a given cosmological model. This step in turn requires building well-understood search algorithms that can be run on artificial dwarf galaxies ingested in the survey data. In the case of the PAndAS survey, \citet{Martin2013} developed a likelihood-based algorithm that runs on the survey's photometric catalogue and determines the probability of there being a dwarf galaxy at any location of the survey based on the distribution of local stars on the sky and in the color-magnitude space. \cite{Doliva2022} then used this algorithm to characterize the detection limits of the survey, ingesting half a million artificial dwarf galaxies with varying sizes, luminosities, and positions. The resulting detection limits show significant variations that are driven, as expected, by the size and luminosity of the systems (\ie their surface brightness), but also by the location in the survey. With PAndAS spanning more than $20\deg$ on the sky and M31 being located fairly close to the Milky Way plane ($b=-22\deg$), the strongly varying MW foreground contamination between the southern edge of the survey ($b\sim-35\deg$) and its northern edge ($b\sim-11\deg$) leads to significant variations of the surface brightness limits (from $\sim$30.5 mag/arcsec$^2$ far from the MW plane to $\sim$29 mag/arcsec$^2$ closest to the plane, respectively). In contrast to MW dwarf galaxy searches, variations in the foreground contamination are at least as important as changes to the heliocentric distance of a satellite. Both lead to variations of $\sim$ 1.5 mag/arcsec$^2$ in the surface brightness detection limit over the M31 halo.

With this knowledge in mind, it is possible to reliably infer the global properties of the dwarf galaxy satellite system of M31, taking detection limit biases into account. Among those global properties, the shape of its luminosity function is an important observational probe as it is sensitive to cosmology, to feedback processes and to reionization. Indeed, the normalization, shape and/or the existence of a break in the faint-end of the luminosity function is an imprint of the properties of dark matter \citep{Spergel2000,Bode2001} and of the suppression of star formation from stellar outflows and reionization \citep{Bullock2000,Somerville2002,Mashchenko2008,Koposov2009,Wheeler2015,Boylan-Kolchin2015,Weisz2017}. The radial distribution of dwarf galaxy satellites around their host can also be shaped by the physics of reionization \citep{Ocvirk2011,Dooley2017} and the disruption of subhaloes by the central disk \citep{DOnghia2010, Kelley2019, Samuel2020}.

Another challenge comes from the distribution of satellites that does not appear as isotropic as expected from $\Lambda$CDM \citep{Pawlowski2018}. A disk-like distribution of satellite dwarf galaxies was found around the MW, M31 and Centaurus A \citep{Lynden1976,Kroupa2005,Metz2007,Conn2013,Ibata2013,Muller2018}. In addition, when looking at the position of M31 dwarf galaxies, most of them appear to lie closer to the MW than on the opposite hemisphere \citep{McConnachie2005b,Conn2012,Wan2020}. With updated RRLyrae-based distances, \cite{Savino2022} reenforces those conclusions, further highlighting the anisotropy in the M31 satellite distribution. The detection limits of dwarf galaxy searches could lead to an anisotropy between the close and far hemispheres of M31 as the farther the dwarf galaxy the harder it is to detect. It is therefore essential to fold in these detection limits when inferring the global properties of the M31 satellite system to check if this anisotropy could simply be the results of observational biases.

Faced with the issue of comparing biased, incomplete observations with models, it may seem more convenient and straightforward to simply correct observed properties, for instance a binned luminosity function, with correction factors calculated from the detection limits. This technique is however plagued by noise in the case of small samples, as is the case for dwarf galaxy systems. Therefore, while it is computationally more expensive, it is much more reliable to forward model the limitations of the data (detection limits, irregular shape of the survey) directly into the model; this is the approach that we follow here.

We use a forward-modeling approach to infer the combined properties of the luminosity function, the radial distribution, and the size-luminosity relation of the dwarf galaxy system of M31. In Section~\ref{Preliminaries}, we detail the sample of observed satellites and the dwarf galaxy completeness of PAndAS. Section~\ref{Basics of the model} and \ref{Likelihood function}, describes the framework and the model used to obtain the results presented in Section~\ref{Results}. Finally, we summarize and discuss the main properties of the dwarf galaxy satellite system of M31 in Section~\ref{Discussion and conclusion}.

\section{Sample} \label{Preliminaries}
PAndAS \citep{McConnachie2009, McConnachie2018} was conducted from 2008 - 2011 with the 1 square degree MegaCam wide field image at the Canada-France-Hawaii Telescope (CFHT). Combined with previous observations \citep{Ibata2007,McConnachie2008}, this Large Program resulted in a survey of over 400 square degrees surrounding M31 and M33, reaching out to $\sim150\kpc$ and $\sim50\kpc$ in projected distance from these galaxies, respectively. For the details of the survey and the creation of the catalogues, we refer the reader to \citet{McConnachie2018}, but it is worth mentioning that the $g$ and $i$ band photometry is obtained for all fields with a median depth of 26.0 and 24.8 for $5\sigma$ detections, respectively \citep{Ibata2014}.

The 24 dwarf galaxies known within this footprint are listed in Table~\ref{sample_dwarf}. The luminosity and size of each dwarf galaxy are taken from \cite{Martin2016} and \cite{Savino2022}. Where needed, distance-related properties (physical half-light radii, absolute magnitudes) are updated using the distances from \cite{Savino2022}. Given the uncertain nature of And~XXVII that may well be a disrupted system \citep{Preston2019} and has large uncertainties in its structural properties \citep{Richardson2011,Martin2016}, we choose not to add it to our sample.

The search for dwarf galaxies suffers from spatial and photometric incompleteness. The former arises from the complex PAndAS coverage on the sky and its correction is quite straightforward, while the latter stems from the complex detection process and is very sensitive to the characteristics of a given dwarf galaxy but also to its location within the survey, mainly because of the varying MW and M31 stellar contamination. The detection limits were derived by \cite{Doliva2022} via the ingestion of nearly half a million artificial dwarf galaxies in the PAndAS catalogue to obtain the recovery fraction for each MegaCam field on a $M_{V}$ and $\log(r_\textrm{h(pc)})$ grid defined by $-8.5\leq M_{V} \leq-4.5$ and $1.8\leq \log(r_\textrm{h(pc)})\leq3.0$, with a step size of 0.25 and 0.10, respectively. The recovery (or lack thereof) of a dwarf galaxy is performed with the search algorithm developed by \citet{Martin2013} and that looks for overdensities of stars both spatially and along a red-giant-branch feature in the color-magnitude diagram. An analytical model is fitted to the resulting $M_{V}$--$\log(r_\textrm{h(pc)})$ recovery fraction grid so that the recovery fraction of any galaxy at a given location, with a given magnitude and size, can easily be calculated. We have also built an analytical model to account for the impact of the heliocentric distance to a dwarf galaxy on recovery fractions. Although the impact of distance to a dwarf galaxy is less important than other parameters, the effect is still not negligible and needs to be taken into account \citep{Doliva2022}. From these, the resulting efficiency of detection for all 24 dwarf galaxies in the sample are listed in Table~\ref{sample_dwarf}.

\begin{table*}
\begin{center}
  \caption{Sample of the dwarf galaxies present in the PAndAS survey.}
\begin{tabular}{llllllllll}
  \hline
  Name &$\alpha$(J2000) & $\delta$(J2000) & $r\textrm{h}$(arcmin) & $m_{V}$ & $D_\mathrm{MW}(kpc)$ & $M_V$& $r\textrm{h}$(\pc) & $D_\mathrm{M31}(\kpc)$ &Recovery fraction \\
  \hline
And~I 		&00$^\textrm{h}$45$^\textrm{m}$39.7$^\textrm{s}$ 	&+38$\deg$02$'$15$''$  	& $3.9_{-0.1} ^{+0.1}$ 	&$13.1_{-0.2} ^{+0.2}$ 	&775$_{-17}^{+19}$	&-11.4$\pm$0.2		&880$^{+31}_{-30}$		&48.0$^{+10}_{-3.2}$	&1.00 \\
And~II 		&01$^\textrm{h}$16$^\textrm{m}$26.8$^\textrm{s}$  	&+33$\deg$26$'$07$''$   	& $5.3_{-0.1}^{+0.1}$  	&$12.4_{-0.2} ^{+0.2}$	&667$_{-15}^{+16}$	&-11.7$\pm$0.2		&1028$^{+31}_{-30}$	&168.9$^{+19}_{-16}$	&1.00\\
And~III   		&00$^\textrm{h}$35$^\textrm{m}$30.9$^\textrm{s}$  	&+36$\deg$29$'$56$''$    	& $2.0_{-0.2} ^{+0.2}$  	&$14.8_{-0.2} ^{+0.2}$	&721$_{-16}^{+17}$	&-9.5$\pm$0.2			&420$\pm$43			&84.9$^{+19}_{-14}$		&1.00    \\
And~V     		&01$^\textrm{h}$10$^\textrm{m}$17.5$^\textrm{s}$  	&+47$\deg$37$'$42$''$    	& $1.6_{-0.1} ^{+0.2}$   	&$15.1_{-0.2} ^{+0.2}$	&759$_{-20}^{+21}$	&-9.3$\pm$0.2			&353$^{+35}_{-24}$		&110.5$^{+7}_{-3.5}$	&1.00  \\
And~IX    		&00$^\textrm{h}$52$^\textrm{m}$53.4$^\textrm{s}$  	&+43$\deg$11$'$57$''$    	& $2.0_{-0.2} ^{+0.2}$   	&$15.6_{-0.3} ^{+0.3}$	&702$_{-20}^{+19}$	&-8.6$\pm$0.3			&408$^{+62}_{-42}$		&82.0$^{+26}_{-24}$		&1.00   \\
And~X     		&01$^\textrm{h}$06$^\textrm{m}$35.4$^\textrm{s}$  	&+44$\deg$48$'$27$''$    	& $1.1_{-0.2} ^{+0.4}$   	&$16.7_{-0.3} ^{+0.3}$	&630$_{-18}^{+18}$	&-7.3$\pm$0.3			&202$^{+74}_{-37}$		&162.2${+25}_{-24}$		&1.00	  \\
And~XI    		&00$^\textrm{h}$46$^\textrm{m}$19.7$^\textrm{s}$ 	&+33$\deg$48$'$10$''$    	& $0.6_{-0.2} ^{+0.2 }$  	&$18.0_{-0.4} ^{+0.4 }$	&751$_{-22}^{+23}$	&-6.4$\pm$0.4			&131$\pm$44			&104.2$^{+11}_{-4.2}$	&0.97	  \\
And~XII   		&00$^\textrm{h}$47$^\textrm{m}$28.3$^\textrm{s}$ 	&+34$\deg$22$'$38$''$    	& $1.8_{-0.7} ^{+0.2}$   	&$17.7_{-0.5} ^{+0.5}$	&718$_{-26}^{+25}$	&-6.6$\pm$0.5			&376$^{+251}_{-147}$	&107.7$^{+20}_{-13}$	&0.80	   \\
And~XIII  		&00$^\textrm{h}$51$^\textrm{m}$51.0$^\textrm{s}$  	&+33$\deg$00$'$16$''$    	& $0.8_{-0.3} ^{+0.4}$   	&$17.8_{-0.4} ^{+0.4}$ 	&821$_{-26}^{+28}$	&-6.8$\pm$0.4			&191$^{+96}_{-72}$		&126.4$^{+16}_{-8.0}$	&0.99  \\
And~XIV   	&00$^\textrm{h}$51$^\textrm{m}$35.0$^\textrm{s}$  	&+29$\deg$41$'$23$''$    	& $1.5_{-0.2} ^{+0.2}$   	&$15.8_{-0.3} ^{+0.3}$	&773$_{-21}^{+21}$	&-8.6$\pm$0.3			&337$\pm$46			&160.8$^{+3.8}_{-4.2}$	&1.00 \\
And~XV    	&01$^\textrm{h}$14$^\textrm{m}$18.3$^\textrm{s}$  	&+38$\deg$07$'$11$''$	& $1.3_{-0.1} ^{+0.1}$   	&$16.0_{-0.3} ^{+0.3}$	&746$_{-18}^{+17}$	&-8.4$\pm$0.3			&283$\pm$23			&95.8$^{+12}_{-4.8}$	&1.00	   \\
And~XVI   	&00$^\textrm{h}$59$^\textrm{m}$30.3$^\textrm{s}$  	&+32$\deg$22$'$34$''$	& $1.0_{-0.1} ^{+0.1}$   	&$16.1_{-0.3} ^{+0.3}$	&517$_{-19}^{+18}$	&-7.5$\pm$0.3			&239$\pm$25			&280.0$^{+26}_{-27}$	&1.00   \\
And~XVII  	&00$^\textrm{h}$37$^\textrm{m}$06.3$^\textrm{s}$  	&+44$\deg$19$'$23$''$	& $1.4_{-0.3} ^{+0.3}$   	&$16.6_{-0.3} ^{+0.3}$	&757$_{-23}^{+24}$	&-7.8$\pm$0.3			&315$\pm$68			&49.9$^{+17}_{-5.8}$	&1.00	  \\
And~XIX   	&00$^\textrm{h}$19$^\textrm{m}$34.5$^\textrm{s}$  	&+35$\deg$02$'$41$''$	& $14.2_{-1.9} ^{+3.4}$ 	&$14.5_{-0.3} ^{+0.3}$	&813$_{-31}^{+31}$	&-10.1$\pm$0.3		&3357$^{+816}_{-465}$	&113.3$^{+18}_{-6.9}$	&1.00 	    \\
And~XX    	&00$^\textrm{h}$07$^\textrm{m}$30.6$^\textrm{s}$ 	&+35$\deg$07$'$37$''$	& $0.4_{-0.1} ^{+0.2}$   	&$18.0_{-0.4} ^{+0.4}$	&741$_{-27}^{+27}$	&-6.4$\pm$0.4			&86$^{+43}_{-22}$		&128.4$^{+12}_{-5.5}$	&0.98\\
And~XXI   	&23$^\textrm{h}$54$^\textrm{m}$47.9$^\textrm{s}$  	&+42$\deg$28$'$14$''$	& $4.1_{-0.4} ^{+0.8}$   	&$15.5_{-0.3} ^{+0.3}$ 	&770$_{-22}^{+23}$	&-8.9$\pm$0.3			&922$^{+182}_{-95}$	&124.4$^{+5.1}_{-3.8}$	&1.00	 \\
And~XXII  	&01$^\textrm{h}$27$^\textrm{m}$40.4$^\textrm{s}$  	&+28$\deg$05$'$25$''$	& $0.9_{-0.2} ^{+0.3}$   	&$18.0_{-0.4} ^{+0.4}$	&754$_{-23}^{+24}$	&-6.4$\pm$0.4			&198	$^{+66}_{-44}$		&216.8$^{+5.7}_{-5.6}$	&0.90 \\
And~XXIII 	&01$^\textrm{h}$29$^\textrm{m}$21.0$^\textrm{s}$  	&+38$\deg$43$'$26$''$	& $5.4_{-0.4} ^{+0.4}$   	&$14.6_{-0.2} ^{+0.2}$  	&745$_{-25}^{+24}$	&-9.8$\pm$0.2			&1170$^{95}_{94}$		&128.1$^{10}_{-4.9}$	&1.00\\
And~XXIV 	&01$^\textrm{h}$18$^\textrm{m}$32.7$^\textrm{s}$  	&+46$\deg$22$'$13$''$	& $2.6_{-0.5} ^{+1}$   	&$16.3_{-0.3} ^{+0.3}$ 	&609$_{-20}^{+19}$	&-7.6$\pm$0.3			&460$^{+178}_{-90}$	&194.5$^{+25}_{-24}$	&0.92	 \\
And~XXV  	&00$^\textrm{h}$30$^\textrm{m}$09.9$^\textrm{s}$  	&+46$\deg$51$'$41$''$	& $2.7_{-0.2} ^{+0.4}$   	&$15.3_{-0.2} ^{+0.3}$ 	&752$_{-23}^{+23}$	&-9.1$^{+0.3}_{-0.2}$	&590$^{+90}_{-47}$		&85.2$^{+12}_{-4.4}$	&1.00	 \\
And~XXVI  	&00$^\textrm{h}$23$^\textrm{m}$46.3$^\textrm{s}$  	&+47$\deg$54$'$43$''$	& $1.0_{-0.5} ^{+0.6}$   	&$18.5_{-0.5} ^{+0.7}$ 	&786$_{-23}^{+24}$	&-6.0$^{+0.7}_{-0.5}$	&229$^{+138}_{-115}$	&104.6$^{+6.8}_{-3.5}$	& $7.00 \times 10^{-5}$ \footnote{The very low detection rate for And~XXVI could imply that it is more massive/luminous than its uncertain measurement.}  \\
And~XXX   	&00$^\textrm{h}$36$^\textrm{m}$34.6$^\textrm{s}$  	&+49$\deg$38$'$49$''$	& $1.5_{-0.2} ^{+0.2}$   	&$16.0_{-0.2} ^{+0.3}$ 	&558$_{-16}^{+17}$	&-7.7$^{+0.3}_{-0.2}$	&245$\pm$33			&238.6$^{+24}_{-24}$	&1.00  \\
NGC~147   	&00$^\textrm{h}$47$^\textrm{m}$27.0$^\textrm{s}$  &+34$\deg$22$'$29$''$	& $6.70\pm0.09$   		&$7.76\pm0.06$ 		&773$_{-20}^{+21}$	&-16.6$\pm$0.07		&1431$^{+44}_{-43}$	&107.0$^{+15}_{-8}	$	&1.00\\
NGC~185   	&00$^\textrm{h}$38$^\textrm{m}$58.0$^\textrm{s}$  &+48$\deg$20$'$15$''$	& $2.94\pm0.04$   		&$8.46\pm0.06$ 		&650$_{-18}^{+18}$	&-15.6$\pm$0.07		&555$\pm$17			&154.1$^{+23}_{-21}$	&1.00	\\
  \end{tabular}
  \end{center}
\textbf{Notes:} The apparent magnitude and apparent size values are taken from \cite{Martin2016}, except for those of NGC 147 and NGC 185 that are taken from \cite{Crnojevic2014}. All absolute magnitudes, physical sizes and distances are from \cite{Savino2022}. While being in the PAndAS footprint, some galaxies are not part of this sample because they are in a region where the completeness was not determined (M32, NGC205; \citealt{Doliva2022}), because their structural parameters are too uncertain (And~XXVII; \citealt{Richardson2011}), or because their distances from M31 is beyond $300\kpc$ (And~XVIII; \citealt{Savino2022}).
  \label{sample_dwarf}
\end{table*}

\section{Model} \label{modelsection}
Here, we discuss our methodology to infer the global properties of the M31 dwarf galaxy system from the observed properties of the individual M31 dwarf galaxies. Accounting for the PAndAS survey detection limits, we infer the underlying M31 luminosity function, size-luminosity relation, and spatial distribution via forward modelling.

\subsection{Dwarf galaxy probabilistic model}  \label{Basics of the model}

Consider a dwarf galaxy whose observed properties, listed in Table~\ref{sample_dwarf}, are: its coordinates on the sky, $(\alpha,\delta)$, its apparent magnitudes in both the $g$ and $i$ PAndAS bands, $m_\textrm{g}$ and $m_\textrm{i}$, its angular half-light radius, $r_\textrm{h}^\mathrm{ang}$, and its heliocentric distance, $D_\textrm{MW}$. As detailed in \cite{McConnachie2018}, the apparent magnitude are corrected for extinction following \cite{Schlegel1998} and \cite{Schlafly2011} and $m_\textrm{V}$ is obtained from $m_\textrm{g}$ and $m_\textrm{i}$ using the transformation equations derived in \cite{Ibata2014}. Then, using $D_\textrm{MW}$, it is straightforward to transform these observed properties into the intrinsic properties of the systems: the absolute magnitude, $M_\textrm{V}$, and the physical half-light radius, $r_\textrm{h}$. We also use the observed properties of the dwarf galaxy $D_\textrm{MW}$ and ($\alpha,\delta$) to calculate the spherical coordinates of a dwarf galaxy in the M31-centric referential $(r_\textrm{M31}, \theta, \phi)$. The dwarf galaxy properties considered here are therefore $\mathcal{D}=\{M_\mathrm{V}, \log r_\textrm{h}, r_\textrm{M31}, \theta, \phi\}$.

We chose to define the dwarf galaxy probabilistic model that depends on a set of parameters $\mathcal{P}$ as the combination of three independent components: the probability of a dwarf galaxy to have a given absolute magnitude,\ie the shape of the luminosity function of the satellite system, $P_\mathrm{LF}(M_\mathrm{V}|\mathcal{P})$; the probability for a dwarf galaxy to have a given size knowing its magnitude,\ie the size-luminosity relation of the satellite system, $P_{\log r_{\textrm{h}}|M_\textrm{V}}(\log r_\textrm{h} | M_\textrm{V},\mathcal{P})$; and the probability of a dwarf galaxy to be at a given (sky-projected or 3-dimensional) location around M31, $P_\mathrm{sp}(r_\textrm{M31},\theta, \phi| \mathcal{P})$. We assume that those components are independent \footnote{ While we could expect galaxies that orbit close to the central parts of the M31 halo to be more easily destroyed by tidal effects, the orbits of most M31 satellites are currently unknown and this process is difficult to take into account.} of each other, which allows us to simply write the probabilistic model as 
\begin{equation}
\label{eq:model}
\begin{split}
P(\mathcal{D}| \mathcal{P})\propto& P_\mathrm{LF}(M_\textrm{V}  | \mathcal{P})\,P_{\log r_{\textrm{h}}|M_\textrm{V} }(\log r_\textrm{h} | M_\textrm{V},\mathcal{P})\\
&P_\mathrm{sp}(r_\textrm{M31}| \mathcal{P}).
\end{split}
\end{equation}

Following \cite{Tollerud2008}, we model the shape of the luminosity function of the satellite system as a power law with exponent $\beta$ over the magnitude range that we consider here for M31 dwarf galaxies ($M_V<-5.5$)\footnote{It is expected that the luminosity function of dwarf galaxies is, at the faint end, truncated by physical processes \citep[e.g.,][]{Bullock2017,Simon2019}. While reionization is expected to affect galaxies as bright as $M_\textrm{V}$=7.0 \citep{Brown2014, Weisz2014}, there are not enough faint M31 satellites to robustly constrain a more complex model that would include this effect \citep[\eg][]{Koposov2009}. We therefore choose to extrapolate our model until $M_\textrm{V}=-5.5$, which might results in a slightly optimistic number of dwarf galaxies expected around M31.}:

\begin{equation}
P_{LF}(M_\textrm{V}|\beta)\propto\frac{\log 10}{2.5}10^{-(\beta+1) (M_\textrm{V}-4.83)/2.5}.
\end{equation}

Following \cite{Shen2003} and \cite{Brasseur2011}, we assume a linear relation between $M_\textrm{V}$ and the mean  $\log(r_\textrm{h})$, $\langle\log r_\textrm{h}\rangle$, such that
\begin{equation}
\langle\log r_\textrm{h}\rangle=z_p+s (M_\textrm{V}+6.0),
\end{equation}
with $s$ the slope and $z_p$ the value of the relation for $M_V=-6.0$. The intrinsic dispersion, $\sigma$, around the relation is modeled as a Gaussian distribution along the $\log r_\textrm{h}$ direction and yields

\begin{equation}
\begin{split}
P_{\log r_{\textrm{h}}|M_\textrm{V}}&(\log r_\textrm{h} | M_\textrm{V},z_p,s,\sigma)=\\
&\frac{1}{\sqrt{2\pi}\sigma} \exp\left(-\frac{1}{2}\left(\frac{\log r_\textrm{h}-\langle\log r_\textrm{h}\rangle}{\sigma}\right)^2\right).
\end{split}
\end{equation}

Finally, as the distribution of M31 satellite dwarf galaxies appears to be circularly but not spherically isotropic \citep{Savino2022}, we consider two cases for the spatial distribution part of the model: a sky-projected (2D) and a volumic (3D) distribution model. In both cases, we choose a simple and agnostic shape for the radial density distribution function, a power law, with parameters $\alpha_\textrm{2D}$ and $\alpha_\textrm{3D}$, respectively\footnote{It may be tempting to assume models informed by the distribution of dark matter sub-halos in simulations, such as, for example, an NFW profile \citep{Navarro1996}. However, considering that the region with $r_\textrm{M31}<30$ kpc is masked in our study, the concentration of the profile would be difficult to constrain. We will further explore a more complex modeling of the radial density distribution function in a futur contribution.}. At this stage, we introduce the assumption of an isotropic distribution of the dwarf galaxies around M31 (an assumption we will revisit later) to simplify the problem at hand. This allows us to remove the impact of the spherical coordinate angles on any model we define and, for the 3D case, we have

\begin{equation}
\begin{split}
P_\mathrm{sp}(r_{M31}, \theta, \phi | \alpha_{3D}) &= P_\mathrm{sp}(r_{M31} | \alpha_{3D})\\
&\propto\int_0^{2\pi}\int_0^\pi r^2 r^{\alpha_{3D}} \sin(\theta) d\theta d\phi \\
&\propto4\pi r^{2+\alpha_{3D}}.\\
\end{split}
\end{equation}

\noindent Similarly, for the sky-projected model, 

\begin{equation}
\begin{split}
P_\mathrm{sp}(r_\textrm{M31} | \alpha_\textrm{2D})&\propto2\pi r^{1+\alpha_\textrm{2D}}.\\
\end{split}
\end{equation}
In summary, the probabilistic model has 5 parameters $\mathcal{P}= \{\beta,z_p,s,\sigma,\alpha\}$, with $\alpha=\alpha_\mathrm{2D}$ or $\alpha=\alpha_\mathrm{3D}$ in the 2D and 3D cases, respectively. Folding everything together and introducing the normalization constant $A(\mathcal{P})$ to ensure that $P(\mathcal{D}| \mathcal{P})$ integrates to unity, Equation \ref{eq:model} becomes

\begin{equation}
\begin{aligned}
P(\mathcal{D}| \mathcal{P}) = &A( \mathcal{P})\,P_\mathrm{sp}(r_\textrm{M31} | \alpha_\textrm{3D})\,P_{\log r_{\textrm{h}}|M_\textrm{V}}(\log r_\textrm{h} | M_\textrm{V},z_p,s,\sigma) \\
&P_{LF}(M_\textrm{V}|\beta) ,\\
\end{aligned}
\end{equation}

\noindent with $A(\mathcal{P})$ such that
\begin{equation}
\iiint P(\mathcal{D}| \mathcal{P}) dr_\textrm{M31} d\log r_\textrm{h} dM_\textrm{V}=1.
\end{equation}

\subsection{Final likelihood function} \label{Likelihood function}

The probabilistic model presented above describes the distribution of dwarf galaxies in the data space but does not provide any constraint on the theoretical number of dwarf galaxies that inhabit the M31 satellite system, $N_\textrm{true}$, over the chosen ranges of observed properties\footnote{In all that follows, we choose the magnitude range $-17<M_\textrm{V}<-5.5$ that, at the bright end, includes the brightest M31 dwarf that is in the survey footprint (NGC~147) and, at the faint end, is fainter than the faintest dwarf galaxy known around M31 (And~XXVI; $M_V=-6^{+0.7}_{-0.5}$). The volume we consider is delimited by $30\kpc<r_\textrm{M31}<300\kpc$, bound by a rough estimate of the virial radius of M31 and an inner boundary that corresponds to a region in which the search for dwarf galaxy is made extremely difficult by the presence of the galaxy's disk \citep{Doliva2022}. We also choose to explore a size range of $1.8<\log r_{\textrm{h}}<4$ which encompass the size of all known M31 dwarf galaxies (Table~\ref{sample_dwarf}).}. At this stage we also introduce the data variable $N_\textrm{obs}$ that is the number of observed dwarf galaxies in the considered magnitude range and volume\footnote{While we consider a sample constructed for the 24 dwarf galaxies listed in Table~\ref{sample_dwarf}, $N_\textrm{obs}$ may not always be 24 as our drawing from the uncertainties on the parameters of the dwarf galaxies and, in particular, their distance may, in a small number of cases, push a sample dwarf galaxy outside of the studied volume.}. For simplicity, we define $\mathcal{D'} = \mathcal{D} \cup \{N_\textrm{obs}\}$ and $\mathcal{P'} = \mathcal{P} \cup \{N_\textrm{true}\}$.

To constrain $N_\textrm{true}$ using, in particular, $N_\textrm{obs}$, we add another layer to the statistical framework and now consider the theoretical density function, unaffected by the survey footprint and detection limits, $\rho_{\mathrm{true}}(\mathcal{D}|\mathcal{P'})= N_\textrm{true} P(\mathcal{D}| \mathcal{P})$. Folding in the detections limits yields the observed version of this function, $\rho_{\mathrm{obs,th}}$, simply defined as

\begin{equation}
\begin{split}
\rho_{\mathrm{obs,th}}(\mathcal{D}|\mathcal{P'}) & = \tau(\mathcal{D})\,\rho_{\mathrm{true}}(\mathcal{D}|\mathcal{P'})\\
& = \tau(\mathcal{D})\,N_\textrm{true}\,P(\mathcal{D}| \mathcal{P}),
\end{split}
\end{equation}

\noindent with $\tau(\mathcal{D})$ the probability of detecting a dwarf galaxy depending on its properties $\mathcal{D}$ \citep{Doliva2022}.

Using the formalism of \citet{Kepner1999}, \citet{Rykoff2012} and \citet{Drlica2020_1}, $N_\textrm{true}$ can be constrained by first virtually binning the data space. In any bin $i$, the likelihood $\ell_i(\mathcal{D'}|\mathcal{P'})$ of generating a sample of $N_\textrm{obs,i}$ dwarf galaxy in bin $i$ can be described by the Poisson likelihood  $\mathfrak{P}( N_\textrm{obs,i}\vert N_\textrm{obs,th,i})$. Here, the expectation $N_{\textrm{obs,th},i}$ is the theoretically observed number of dwarf galaxies in bin $i$, or

\begin{equation}
N_{\textrm{obs,th},i}=\tau(\mathcal{D}_i)\,N_\textrm{true}\,P(\mathcal{D}_i| \mathcal{P})\,d\mathcal{D}.
\end{equation}
The total likelihood of the dwarf galaxy system can therefore be expressed as
\begin{equation}
\begin{split}
\mathcal{L}(\mathcal{D'}|\mathcal{P'})&=\displaystyle{\prod_{i\in \mathrm{bins}}}\ell_i(\mathcal{D'}|\mathcal{P'})\\
&=\displaystyle{\prod_{i\in \mathrm{bins}}}\mathfrak{P}( N_\textrm{obs,i}\vert N_\textrm{obs,th,i}) \\
&=\displaystyle{\prod_{i\in \mathrm{bins}}} N_{\textrm{obs,th,}i}^{N_{\textrm{obs,}i}}\exp(-N_{\textrm{obs,th,}i})/N_{\textrm{obs,}i}!, \\
\end{split}
\end{equation}
which further translates into 
\begin{equation}
\label{sumeq}
\footnotesize
\log(\mathcal{L}(\mathcal{D'}|\mathcal{P'}))=-\displaystyle{\sum_{i\in \mathrm{bins}}} N_{\textrm{obs,th,}i} + \displaystyle{\sum_{i\in \mathrm{bins}}} N_{\textrm{obs,}i} \log(N_{\textrm{obs,th,}i})+\textrm{const}.
\end{equation}

The first term of this equation is simply the integral of $\rho_\mathrm{obs,th}$ over the data space. In addition, if we consider bins that are small enough to contain either one or no galaxy, the second part of the equation then becomes a sum over the $N_{\textrm{obs}}$ bins that contain a galaxy. Equation~\ref{sumeq} therefore becomes

\begin{eqnarray}
\label{eq:almost_there}
\log(\mathcal{L}(\mathcal{D'}|\mathcal{P'})) & = & -\displaystyle{\int_\mathcal{D}} \rho_{\textrm{obs,th}} d\mathcal{D} + \displaystyle{\sum_{i=1}^{N_\mathrm{obs}}} \log(N_{\textrm{obs,th,}i})+\textrm{const} \nonumber\\
&= & -\displaystyle{\int_\mathcal{D}} \tau(\mathcal{D}) N_\textrm{true} P(\mathcal{D} | \mathcal{P}) d\mathcal{D}\\
& &+{\sum_{i=1}^{N_\mathrm{obs}}} \log(\tau(\mathcal{D}_i) N_\textrm{true} P(\mathcal{D}_i| \mathcal{P}))+\textrm{const}.\nonumber
\end{eqnarray}

\noindent Here, $\mathcal{D}_i$ are the data values of dwarf galaxy $i$.

With the assumed isotropic distribution of the satellites, the integral of equation~\ref{eq:almost_there} can be marginalized over $\theta$ (in the 2D case) or $\theta$ and $\phi$ (in the 3D case), which introduces the mean fraction of detected dwarf galaxies at radius $r_{\textrm{M31}}$, $\langle\tau(M_V, \log r_\mathrm{h}, r_\textrm{M31})\rangle$. This allows us to drop the dependence on $\theta$ and $\phi$ and the likelihood finally becomes

\begin{eqnarray}
\label{likelihood}
\log(\mathcal{L}(\mathcal{D'}|\mathcal{P'})) & = & -\int_{M_V}\int_{\log\mathrm{r}_{\mathrm{h}}}\int_{r_\mathrm{M31}} \langle\tau(M_V, \log r_\mathrm{h}, r_\textrm{M31})\rangle\nonumber\\
& & N_\textrm{true} P(\mathcal{D} | \mathcal{P})  dr_\mathrm{M31} d\log r_\mathrm{h} dM_V\\
& & + \displaystyle{\sum_{i=1}^{N_\mathrm{obs}}} \log(\tau(\mathcal{D}_i) N_\textrm{true} P(\mathcal{D}_i| \mathcal{P}))+\textrm{const}.\nonumber
\end{eqnarray}

\subsection{Implementation}
We sample the likelihood with our own Metropolis-Hastings algorithm \citep{Metropolis1953,Hastings1970}. In order to obtain the probability distribution function (PDF) for each parameter while taking into account the uncertainties on the observed properties of the dwarf galaxies, we fold in the PDFs on the observed parameters instead of using a single value for each property. Following \citet{Conn2012}, the likelihood becomes the convolution of the likelihood for a single value (Eq~\ref{likelihood}) with the PDF of each observed property of the satellite system. With $\Omega$ the sample of all possible sets of values $\mathcal{D'}$ and $g(\mathcal{D'})$ being the probability of a given set, the likelihood function becomes
\begin{equation}
\mathcal{L}(\mathcal{D'}_\Omega|\mathcal{P'})=\int_\Omega \mathcal{L}(\mathcal{D'}|\mathcal{P'})g(\mathcal{D'})d\mathcal{D'}.
\end{equation}
This integral is calculated numerically via a Monte-Carlo method and the random drawing of 50 satellite systems generated from the PDFs of $m_\textrm{V}$, $r_\textrm{h}$, $D_{MW}$ and $D_{M31}$ for all galaxies in the sample. The final distribution is the sum of the resulting chains.

Finally, we have only considered likelihoods to this point, but we seek to determine the probability of the model given the data $P(\mathcal{P'}|\mathcal{D'}_\Omega)$. It is linked to the probability of the data given the model $\mathcal{L}(\mathcal{D'}_\Omega|\mathcal{P'})$ via the prior $P(\mathcal{P'})$ such that
\begin{equation}
P(\mathcal{P'}|\mathcal{D'}_\Omega)\propto \mathcal{L}(\mathcal{D'}_\Omega|\mathcal{P'})P(\mathcal{P'}).
\end{equation}

\noindent For simplicity, we choose uniform priors on all parameters but we impose that $0<N_\textrm{true}<1000$, $0<\sigma<1$ and $0<z_p<3$. The other parameters do not have additional constraints.

\section{Results} \label{Results}

\subsection{Inferred global properties of the M31 dwarf galaxy system}
\label{Inferred model}

\begin{table*}
\begin{center}
  \caption{Values for the model parameters in the case of a 2D and 3D spatial distribution of dwarf galaxies.}
\begin{tabular}{ccccccc}
  \hline
  &$\beta$ & $z_p$ & $s$ & $\sigma$ & $\alpha$ & $N_\textrm{true}$ \\
  \hline
2D	&$-1.5\pm0.1$&$2.5^{+0.2}_{-0.1}$&$-0.05^{+0.03}_{-0.02}$&$0.32_{-0.05}^{+0.07}$&$-0.1_{-0.5}^{+0.3}$&$136_{-35}^{+65}$\\\\
3D	&$-1.5\pm0.1$&$2.5^{+0.2}_{-0.1}$&$-0.05^{+0.03}_{-0.02}$&$0.33\pm0.06$&$-1.7_{-0.3}^{+0.4}$&$92_{-26}^{+19}$\\
\hline
  \end{tabular}
    \end{center}
\textbf{Notes:} This table presents the inferred global properties of the M31 satellites system in the case of a 2D and 3D spatial distribution. As defined in Section~\ref{modelsection}, $\beta$ is the slope of the luminosity function, $z_p$ is the zero point and $s$ is the slope of the linear relation of dispersion $\sigma$ between the size and the luminosity of dwarf galaxies, $\alpha$ is the slope of the spatial distribution and $N_\textrm{true}$ is the expected number of M31 satellites.
  \label{param2D3D}
\end{table*}

\begin{figure*}
\begin{center}
\includegraphics[width=\hsize]{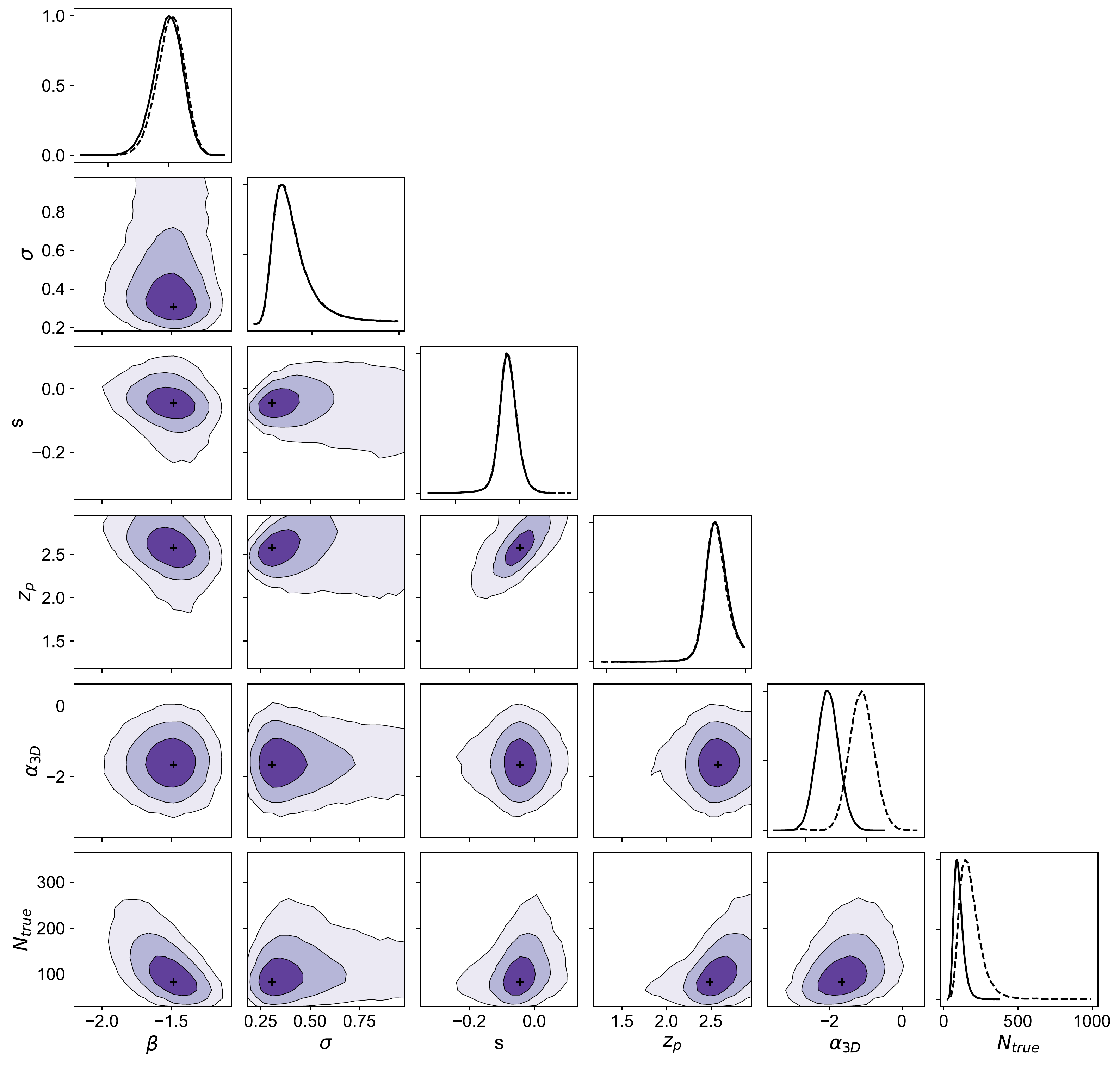}
\caption{\label{corner_plot_3D} Resulting correlation graphs and marginalized PDFs (full line) for each parameter of the 3D model. Black crosses represent the highest likelihood value for each couple of parameters. The marginalized PDFs for the 2D model are represented by the dashed lines.}
\end{center}
\end{figure*}

The contraints on the global dwarf galaxy satellite system of M31 are listed in Table~\ref{param2D3D} for both models with a sky-projected (2D, $\alpha=\alpha_\mathrm{2D}$) and a volumic (3D, $\alpha=\alpha_\mathrm{3D}$) radial distribution component. The marginalized, posterior probability distribution function (PDF) of the different parameters of $\mathcal{P'}$ are presented in Figure~\ref{corner_plot_3D} for the case of the volumic radial distribution model. The favored parameters listed in Table~\ref{param2D3D} correspond to the peak of a parameter's marginalized one-dimensional PDF and the associated credible intervals are bound by the parameter values whose PDF values are 0.61 of the maximum (equivalent to a $\pm1\sigma$ confidence interval in the case of a Gaussian PDF and that we prefer over the 68\% central confidence interval in the case of skewed PDFs).

We first note that all six parameters of the model are well constrained and that the posterior PDFs are rarely perfect Gaussians. This is likely a consequence of the complexity of the model and the non-trivial impact of the detection limits on the model. In addition to constraints on the individual parameters, the statistical framework we have developed makes it very easy to study the correlations (or lack thereof) between different parameters. For instance, we note the strong correlation between $\beta$, the slope of the luminosity function, and $N_\textrm{true}$, the number of M31 dwarf galaxies in the considered volume and magnitude range. This correlation is expected as changes to the slope of the luminosity function will directly lead to a change in the number of dwarf galaxies constrained by the model. Similarly, the correlation between the slope of the size-luminosity relation, $s$, and the value of the slope at $M_V=-6.0$, $z_p$, is expect as there is a tradeoff between making the relation flatter and higher to ensure it goes through the cloud of data points.

\begin{figure}
\begin{center}
\includegraphics[width=\hsize]{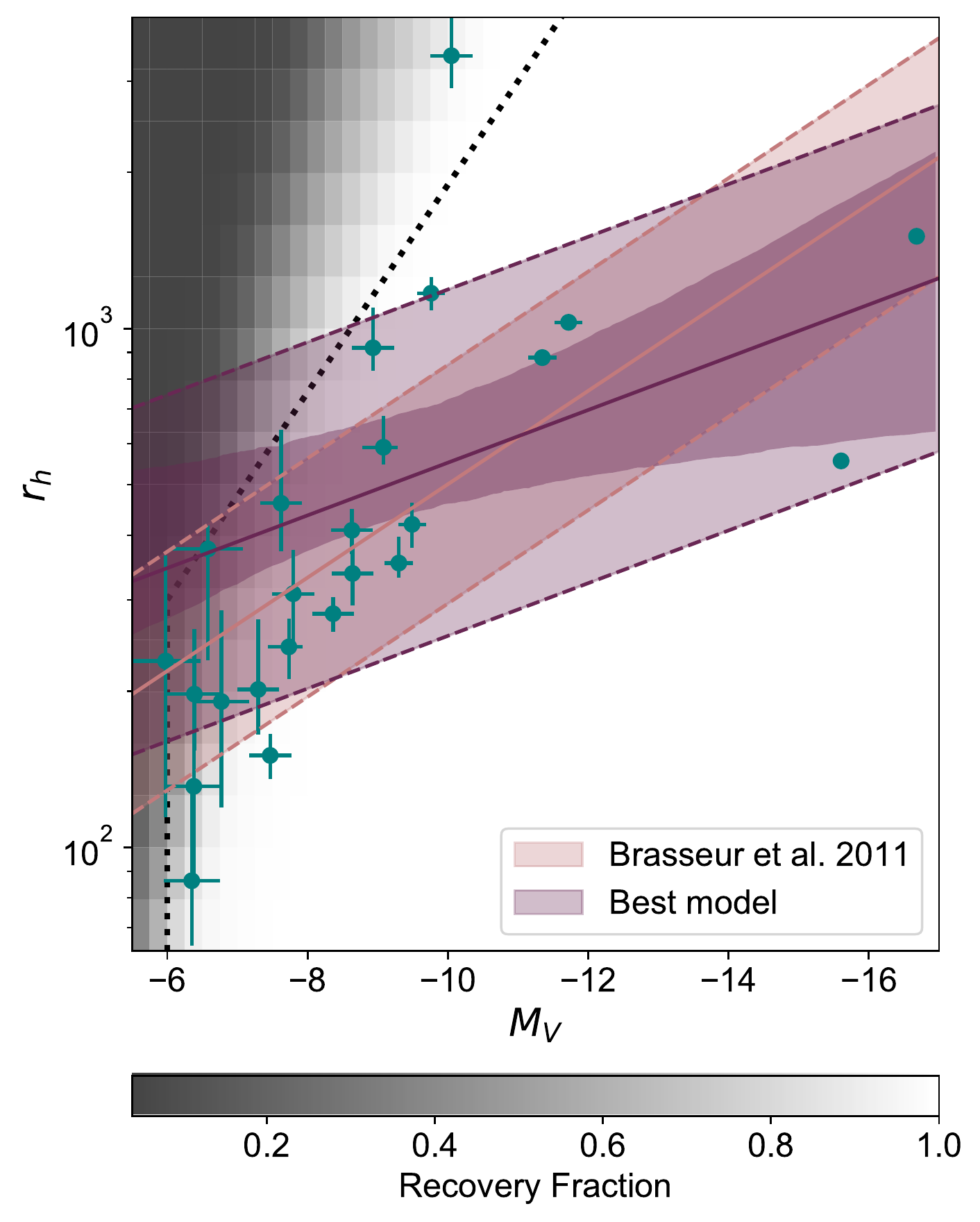}
\caption{\label{mv_rh_3D} Relation between the size and the luminosity of M31's dwarf galaxies as infer through our modeling. The best relation and corresponding width are represented by the purple full line and dashed lines, with the uncertainties on the mean model shown as the high opacity purple band. The best model derived by \cite{Brasseur2011} is represented by the orange lines, and the binary completeness limits they used by the dotted black line. The average detection limits folded in our analysis are represented by the grey background scale (100\% recovery in white and 0\% recovery in dark gray). Given those, the inferred model compensates for the undiscovered large and faint dwarf galaxies and therefore is slightly shifted from what we would na\"ively expect from the cloud of known dwarf galaxies (teal dots).}
\end{center}
\end{figure}

Focusing on this part of the model, the size-luminosity relation for the M31 dwarf galaxies is shown in Figure~\ref{mv_rh_3D}, overlayed on the data of the 24 dwarf galaxies of the sample (teal points with error bars) and the average detection limits determined by \citet{Doliva2022} in this space (gray-scale background). From the marginalized one-dimensional posterior PDFs, we derive $z_p=2.5^{+0.2}_{-0.1}$, $s=-0.05^{+0.03}_{-0.02}$ and $\sigma=0.33\pm0.06$. This relation is similar to but shallower than the one determined by \cite[the light orange model in the figure; $z_p=2.34 \pm 0.1$, $s=-0.09 \pm 0.02$ and $\sigma=0.23_{-0.07}^{+0.02}$]{Brasseur2011}, also determined through forward modeling, but with binary detection limits (recovery fractions of 0 or 1) that follow the dotted line in Figure~\ref{mv_rh_3D}. The differences between the two favored models likely stems from these distinct detection limits and our model infers higher overall values for $r_h$ at the fainter end as it compensates for the large and faint dwarf galaxies that are yet undiscovered because of their low surface brightness limits.

\begin{figure}
\begin{center}
\includegraphics[width=\hsize]{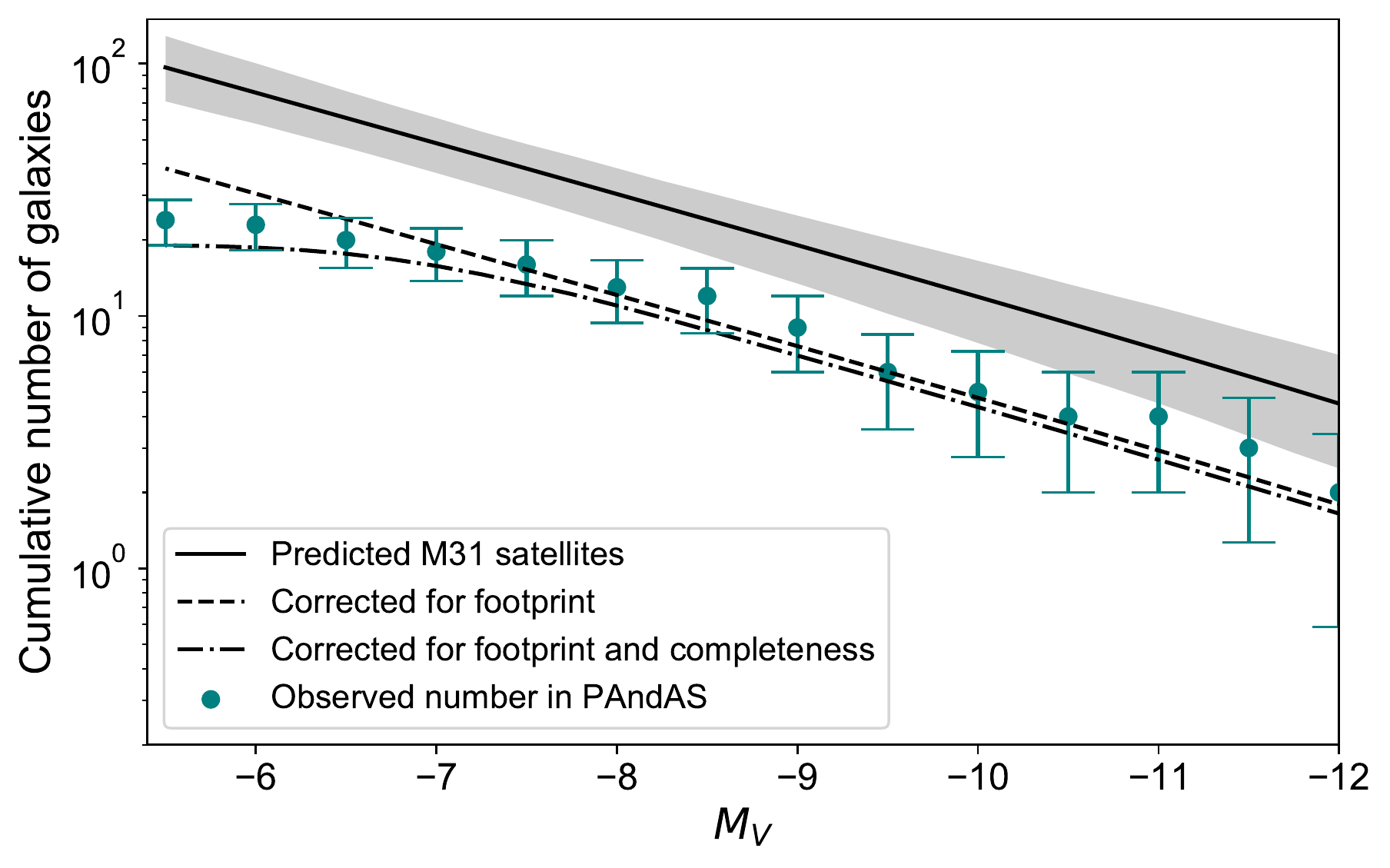}
\caption{\label{luminosity_function_3D} Cumulative number of dwarf galaxies as a function of their magnitude. The inferred model is represented by the black line and gray band. The dashed line shows the inference for the PAndAS footprint and the dot-dashed line the favored model once the average detection limits are applied. This line is directly comparable to, and shows good agreement with, the cumulative distribution known dwarf galaxies (teal points).}
\end{center}
\end{figure}

\begin{figure}
\begin{center}
\includegraphics[width=\hsize]{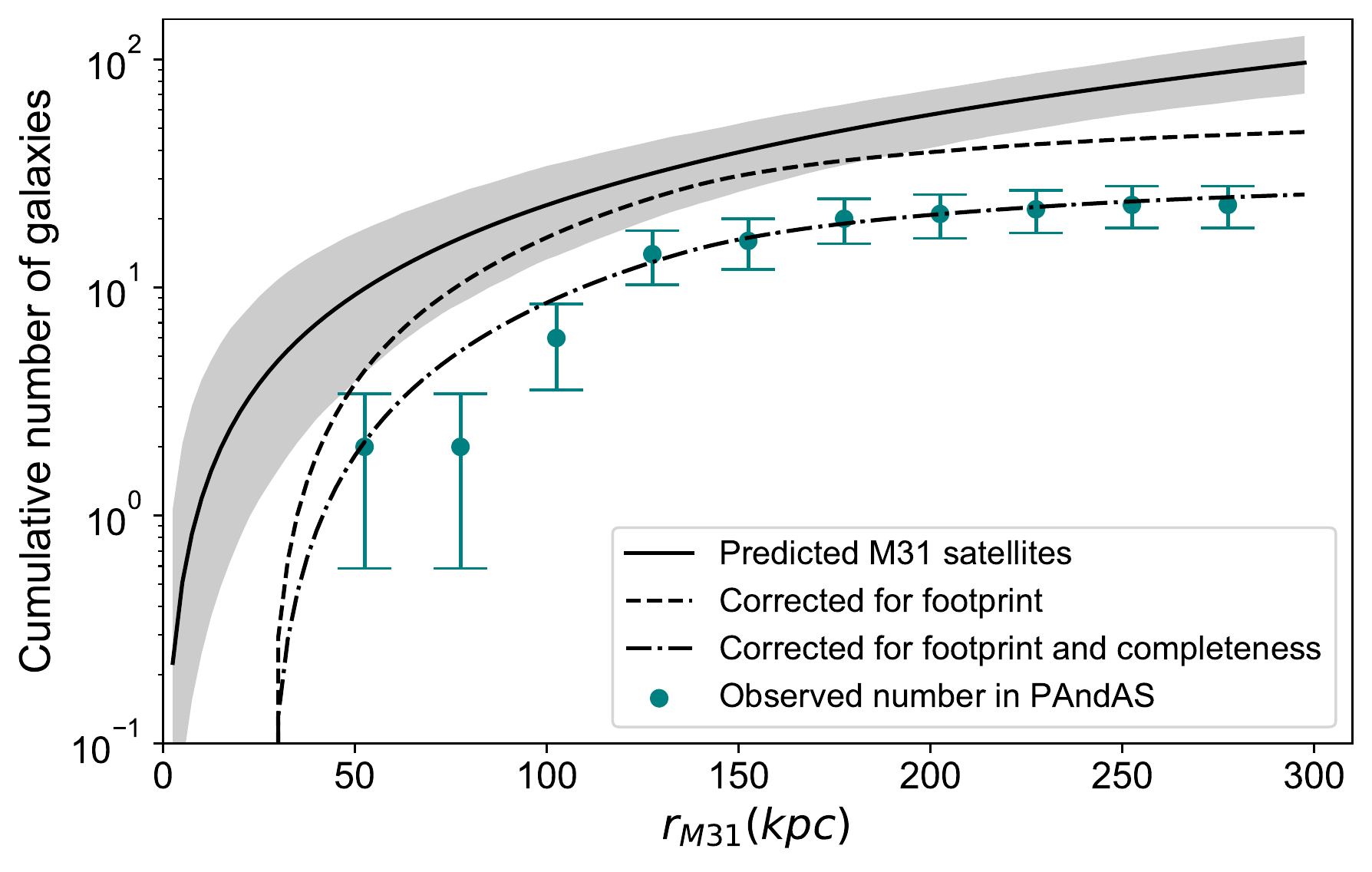}
\caption{\label{distribution_function_3D} Same as Figure~\ref{luminosity_function_3D}, but for the cumulative number of dwarf galaxies as a function of their distance to M31. The favored model, once the spatial and detection limits are accounted for, is compatible with the cumulative distribution known dwarf galaxies.}
\end{center}
\end{figure}

The radial distribution of the assumed isotropic distribution of dwarf galaxies around M31 is well-constrained with $\alpha_\textrm{2D}=-0.1_{-0.5}^{+0.3}$ for the 2D case. It implies an almost flat surface density of dwarf galaxies on the sky and confirms previous hints that this is the case \citep{McConnachie2009}. For the 3D case, we determine a slope $\alpha_\textrm{3D}=-1.7_{-0.3}^{+0.4}$. A previous constraint on the slope of the volumic radial distribution function was provided by \citet{Conn2012}, who determined $\alpha_\textrm{3D}=-1.52_{-0.35}^{+0.32}$ from the forward modeling of a very similar sample of dwarf galaxies but different distance values for the dwarf galaxies\footnote{We now use the updated RRLyrae distances determined by \citet{Savino2022} instead of the tip of the red giant branch distances determined by \citet{Conn2012}.} and also without taking the detection limits into account. The two constraints are nevertheless compatible within their uncertainties and yield a fairly steep density profile, even though it is not as steep as the NFW profile in the external part of the halo \citep[$\beta\sim-3;$][]{Navarro1996}. Also, we note that $\alpha_\textrm{3D}$ is not quite $\alpha_\textrm{2D}-1$, which hints that the surface density model is not simply the integration of the volumic density model and may hint that the chosen models are not a perfect representation of the data. In particular, the assumption of isotropy may not be entirely valid. We revisit this question in the next sub-section.

One of the parameters that is most affected by the detection limits is certainly the slope of the luminosity function as dwarf galaxies that are missed because they are too faint to be discovered in PAndAS will cause the observed luminosity function to drop significantly at faint magnitude. Our analysis yields a strong constraint on this slope and we infer that the intrinsic luminosity function of M31 dwarf galaxies has a slope $\beta = -1.5\pm0.1$ (in the 3D case, similar in the 2D case). It is steeper than the one derived by \citet[$\beta=-1.22_{-0.10}^{+0.11}$]{Crnojevic2019}, who did not model the impact of the detection limits. This difference arises from the consideration of the detection limits of the survey and therefore highlights their importance in deriving the faint end luminosity function.

Finally, we infer the total number of M31 dwarf galaxies within the considered magnitude range ($-17.0<M_V<-5.5$) and volume ($30<r_\mathrm{M31/\kpc}<300$), $N_\mathrm{true}=136_{-35}^{+65}$ in the 2D case, or $N_\mathrm{true}=92_{-26}^{+19}$ in the 3D case. Combined with our framework that only considers galaxies brighter than $M_V=-5.5$ ($\sim 10^{4}\lsun$), and even though the realm of dwarf galaxies extends to much fainter systems (e.g., around the MW; \citealp{McConnachie2012}) that are not detectable in PAndAS \citep{Martin2013}, these values are in line with the expectation that a galaxy like M31 is surrounded by hundreds of dwarf galaxies, most of them faint \citep[\eg][]{Garrison-Kimmel2019}.

To check the quality of the model inference, a comparison of the favored model with the cumulative distribution of observed dwarf galaxies is shown in Figure~\ref{luminosity_function_3D} for the luminosity function and in Figure~\ref{distribution_function_3D} for the volumic radial distribution. The inferred model is shown in black, with the gray band tracking the corresponding uncertainties, while the dashed lines represent the model, corrected to include only the PAndAS footprint, and the dash-dotted line further adds the impact of the detection limits. This final line is directly comparable with the observations (teal points) and shows a good agreement for both cumulative distributions. This is the sign that, despite its complexity, the favored model is a good representation of the known population of M31 dwarf galaxies. These figures also make it evident that the majority of still undiscovered M31 dwarf galaxies brighter than $M_V=-5.5$ are located outside of the PAndAS footprint, mainly beyond $150\kpc$ (the difference between the full and dashed line) but that about half of the dwarf galaxies in the magnitude range $-5.5>M_V>-7.0$ remain to be discovered in the PAndAS footprint (difference between the dashed and dash-dotted line). Some of these are likely to be among the list of candidate satellites already published \citep{Martin2013,Mackey2019}.

\subsection{Anisotropy in the satellite distribution}
\label{anisotropy}

\begin{figure*}
\begin{center}
\begin{minipage}{0.48\hsize}
\includegraphics[width=\hsize]{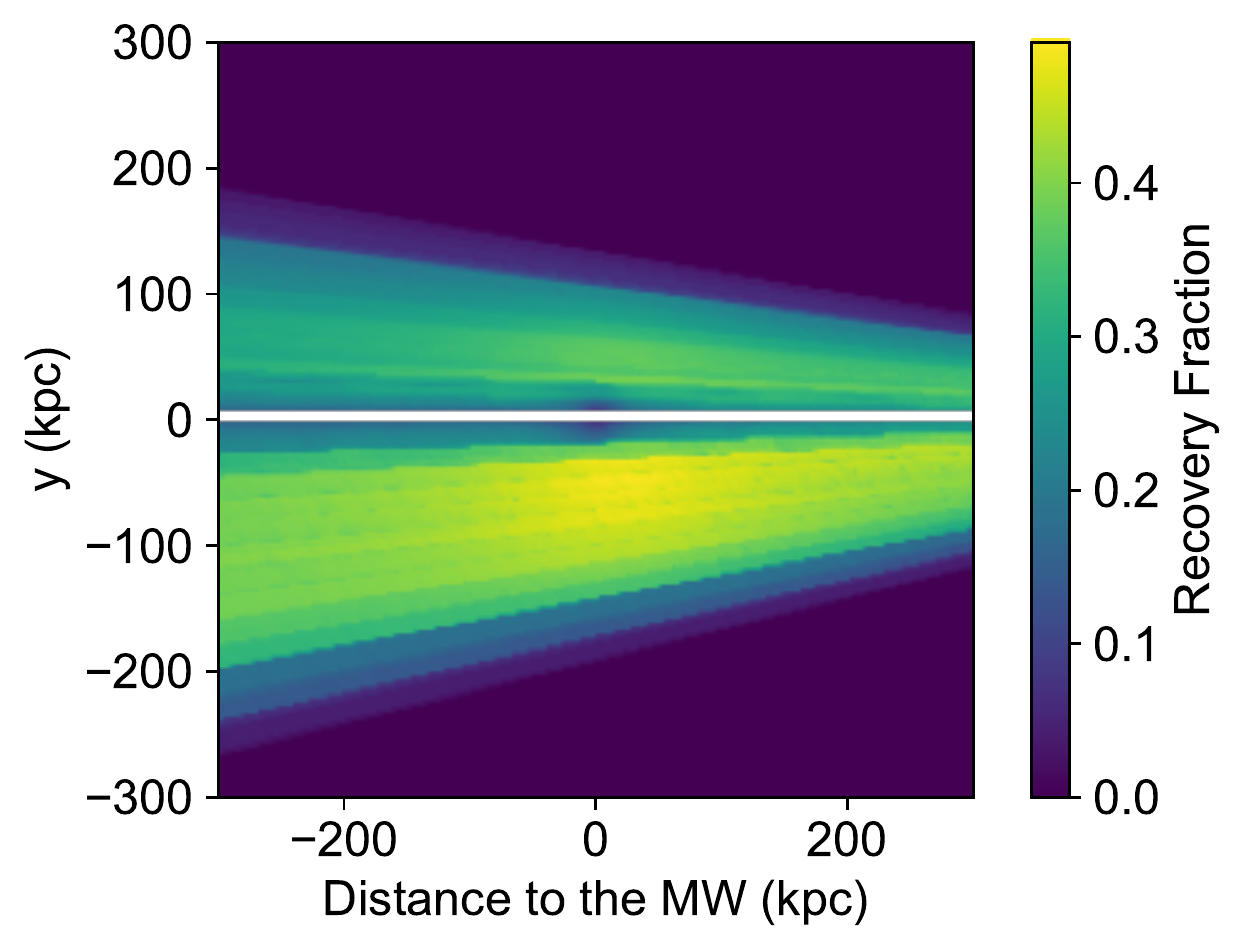}
\end{minipage}
\begin{minipage}{0.48\hsize}
\includegraphics[width=\hsize]{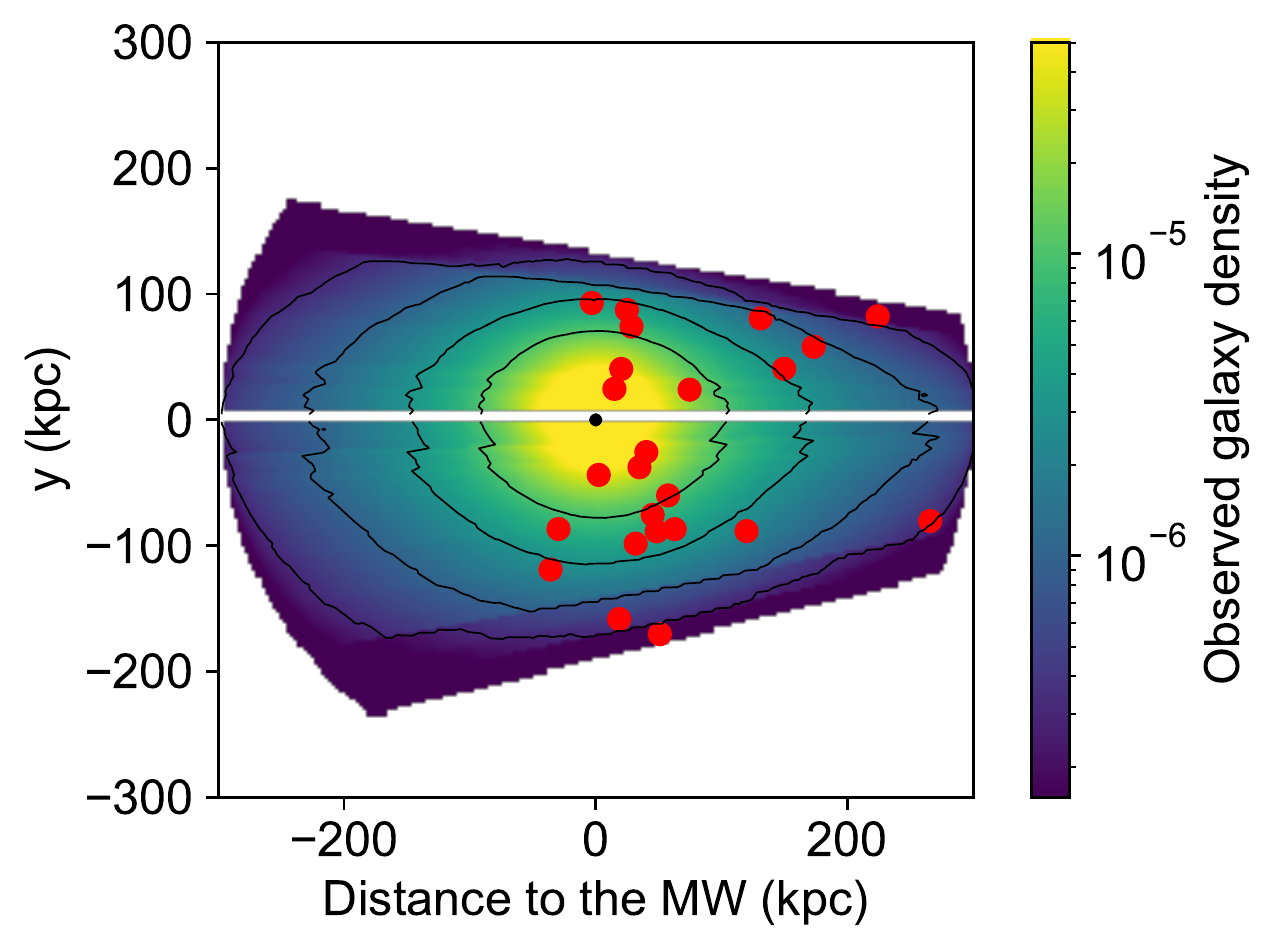}
\end{minipage}
\caption{\label{fig:anisotropy} \textit{Right panel}: Map of the completeness in the y-z plane with z being the distance to the MW. The contamination due to MW and M31 stars lead to the variation of the completeness along the y-axis, while the impact of the distance is visible along the z-axis. \textit{Left panel}: Map of the observed galaxy density in the y-z plane. The red dots represent the known dwarf galaxies. The density is slightly asymmetric as it is somewhat higher on the positive end of the z-axis but this impact might be compensated by the increase of the observed space on the negative end.}
\end{center}
\end{figure*}

With the inferred properties of the isotropic model we have constructed to represent the M31 dwarf galaxy system, we can now explore the perceived anisotropy of the satellite system and, in particular, whether it could be an artefact produced by the detection limits of the PAndAS survey. Looking at the RR Lyrae distances obtained by \cite{Savino2022}, overlaid on the average detection limits in Figure~\ref{fig:anisotropy}, it is clear that the distribution of Andromeda's dwarf galaxies is not isotropic. Among the 24 dwarf galaxies present in the PAndAS footprint and that contribute to our sample, 21 systems are located on the MW side of M31 and produce a strong anisotropy. The (in)completeness impacts the distribution of known satellites in two different ways: it is easier to detect a dwarf galaxy that has a smaller heliocentric distance but, because of the increasing foreground contamination, it is harder to detect a dwarf galaxy closer to the MW plane. Therefore, we aim to test if the favored inferred model, observed through the detection limits (the contours in Figure~\ref{fig:anisotropy}) could naturally produce this observed anisotropy.

To quantify the significance of the anisotropy, we use a simple Monte Carlo procedure to generate 10,000 satellite systems drawn from the favored isotropic model, folding in the PAndAS recovery fractions. In practice, we start by drawing the distance to M31 from the PDF obtained by \cite{Savino2022}. Then, we randomly locate this satellite around M31 using the favored density model and, finally, we test them against the detection limits of \citet{Doliva2022}. We reject dwarf galaxies that do not pass this test and repeat this procedure until the sample of `observed' dwarf galaxies contains 24 satellites.

Of these 10,000 systems drawn from the favored isotropic model, we find that only 7 systems have a distribution that is at least as anisotropic as M31's (at least 21 dwarf galaxies on the MW side of M31). Therefore, we conclude that the asymmetric dwarf galaxy completeness limits of the survey are very unlikely to explain, on their own, the observed anisotropy of the M31 dwarf galaxy satellite system.

\section{Discussion and conclusion} \label{Discussion and conclusion}

\begin{figure}
\begin{center}
\includegraphics[width=\hsize]{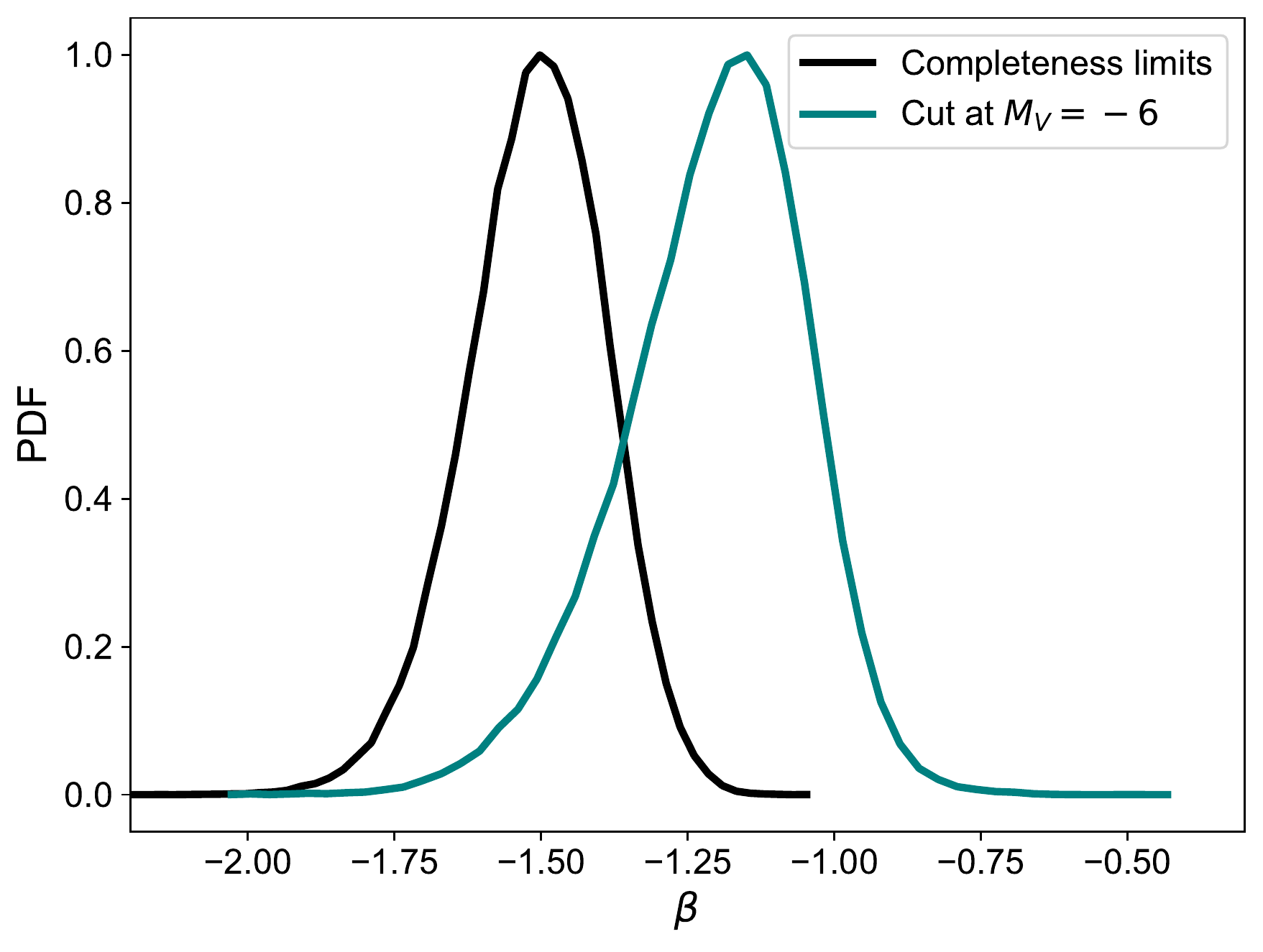}
\caption{\label{completeness_cut} Marginalized probability distribution function of the parameter $\beta$, the slope of the luminosity function, in a case where the dwarf galaxy detection limits are taken into account (black PDF) or they are approximated by a simple magnitude cut (teal PDF).}
\end{center}
\end{figure}

In this paper we inferred the global properties of the M31 dwarf galaxy satellite system within the PAndAS survey through forward modeling, carefully folding in the detection limits of dwarf galaxies in this survey. We conclude that M31 hosts $92_{-26}^{+19}$ or $136_{-35}^{+65}$ satellites, depending on whether the spatial distribution of satellites is modelled in 2D or 3D, over the magnitude range $-17<M_\textrm{V}<-5.5$ and $30<r_\mathrm{M31/\kpc}<300\kpc$, even though only 24 are known within the footprint and 33 overall. Both results would be slightly overestimated if the true luminosity function of M31 satellite dwarf galaxies were to differ from a power law at fainter magnitudes than $M_\textrm{V}=-7$, for instance because of reionization \citep{Brown2014,Weisz2014}. In order to better constrain the luminosity function at the faint end and to model the possible turn over, it is essential to increase the size of the faint dwarf galaxy sample. As shown in Figure~\ref{luminosity_function_3D}, about half of the dwarf galaxies remain undiscovered in the PAndAS footprint and for the chosen magnitude limit. Yet, the most promising regions to search for these still fairly bright but elusive galaxies is to search for them outside the survey footprint: about half of the expected tally of dwarf galaxies brighter than $M_V=-5.5$ reside beyond the edges of the survey, out to the projected virial radius of M31. This also shows the importance of probing a wider area of the M31 surroundings to better constrain its global satellite properties. 
For the MW, \cite{Drlica2020_1} derived a number of $\sim$ 30 satellites with $M_V>$-5.5 and $r_\textrm{M31}<$300 kpc while for M31 we found $92_{-26}^{+19}$ which means that it has $\sim2-3$ times more satellites than the MW until $M_V>$-5.5. This result is in agreement with the number of globular clusters \citep{Huxor2014}  and could easily be explained if M31 is significantly more massive than the MW \citep{Patel2022}.

Assuming a linear relation between $\log r_\textrm{h}$ and $M_V$, we infer a slope $s=-0.05^{+0.03}_{-0.02}$, a zero point $z_p=2.5^{+0.2}_{-0.1}$ at magnitude $M_V=-6.0$, and a scatter $\sigma=0.33\pm0.06$. These are consistent with the values derived by \citet[$z_p=2.38^{+0.16}_{-0.13}$, $s=-0.03\pm 0.03$ and $\sigma=0.2^{+0.08}_{-0.02}$,]{Brasseur2011} for a similar model for the dwarf galaxies of M31 and, also of the MW. We therefore conclude that the satellite population of M31 and the MW do not show any significant difference in their size-luminosity relation.

We assumed that the slope of the luminosity function could be represented by a power law, which we constrained to be $\beta=-1.5\pm0.1$. This result is compatible with previous studies and for other satellite systems. For the MW, \cite{Tollerud2008} found $\beta=-1.9\pm 0.2$ and, for M81, \cite{Chiboucas2013} derived a $\beta=-1.27\pm 0.04$. Tentatively, it would appear that the steepness of the luminosity function may be correlated with the age of the last merger of the host galaxy (about 10\,Gyr for the MW \footnote{While a merger with the LMC satellites and the MW is ongoing \citep{Battaglia2022}, those dwarf galaxies were note taken into account by \cite{Tollerud2008} who studied the satellites in the northern hemisphere.}, 2--4\,Gyr for M31, and ongoing for M81; \eg \citealp{Helmi2018}, \citealp{DSouza2018}, \citealp{Okamoto2019}). It is however important to note that these constraints on $\beta$ were obtained from very different sample sizes and techniques and without always taking a survey's detection limits into account. In the case of M81, for instance, \citealt{Chiboucas2013}, were limited to only 5 galaxies with $M_V> -10$, among which only one was fainter than $M_V=-8.0$. The same authors also derived $\beta=-1.13\pm0.06$ for M31 \citealt{Chiboucas2009}, which is significantly different from our inference, albeit from an earlier sample of known M31 dwarf galaxies and without taking the survey footprint and the detection limits into account.

This comparison underlines the importance of the impact of the detection limits on constraining the global properties of a satellite system. Determining these limits can admittedly be quite a tedious and computationally taxing task \citep{Koposov2008,Drlica2020_1,Doliva2022} but it is absolutely essential to fold them into the analysis. The simplicity of using binary detection limits with magnitude cuts \citep[e.g.,][]{Brasseur2011,Bennet2019,Crnojevic2019} can be appealing but requires a very conservative approach (and throwing away some of the data) to not induce a bias on the slope of the luminosity function. We explore this effect in Figure~\ref{completeness_cut} as we simplify the \citet{Doliva2022} detection limits to binary limits with a magnitude cut at $M_V=-6.0$. This leads to a clear bias on the slope of the luminosity function as undetected galaxies at the faint end are not compensated for. In fact, the value of $\beta$ inferred in this case is similar to those determined by \citet{Chiboucas2009} and \citet{Crnojevic2019} without taking the detection limits into account. The importance of determining accurate detection limits is complementary to the use of bayesian inference. Forward modeling has been used to go from the luminosity function to the stellar-halo mass relation and \textit{vice versa} \citep{Danieli2022}, but the most common method to derive the luminosity function from observations is through their noisy correction from the incompleteness and, in some case, the unadvised fitting of the cumulative function \citep{Chiboucas2009,Geha2017,Crnojevic2019,Bennet2019}. While this avoids the complex work of forward modeling, biases may arise from the loss of information implied by the use of a correction factor(\eg an average of the detection limits, a magnitude cut). And, while fitting the cumulative function may seem like a straight forward process, those data points are intrinsically correlated, which makes it difficult to properly handle statistics and, consequently, uncertainties. The possible correlation between the slope of the luminosity function and the age of the last merger suffered by the host should therefore be consolidated by redetermining the slopes in a homogeneous analysis that would take the completeness of the different surveys into, for example by following the path traced in this paper.

Finally, we show that the radial distribution of the satellites can be modeled by a power law of exponent $\alpha_\textrm{2D}=-0.1_{-0.5}^{+0.3}$ and $\alpha_\textrm{3D}=-1.7_{-0.3}^{+0.4}$ in the sky-projected and volumic radial distribution, respectively. However, we determined that the observed anisotropy of the satellite dwarf galaxies \citep{McConnachie2005b,Savino2022} is unlikely to be the consequence of the detection limits that make it more difficult to detect dwarf galaxies on the more distant side of M31. The observed anisotropy is rarely reproduced by drawing random satellite systems from the favored satellite system model that fold in the limits: only $0.07\%$ of these have an anisotropy that is at least as strong as the observations. Since the incompleteness of the sample is ruled out as a major factor responsible for the observed anisotropy, another observational effect could be a bias in the distances. However, the fact that the anisotropy manifests using both RR Lyrae and TRGB distances disfavor such biases \citep{Conn2012,Savino2022}. In other words, we have, by using robust distances, combined with our detailed modeling of the (in)completeness of the data, built a strong case for the anisotropy to be a real physical configuration. Different solutions are proposed to explain such a distribution \citep{Pawlowski2017,Thomas2018,Wan2020} and complexity could be added to the model presented here to explore these possibilities.

More generically, the statistical framework developed here is very flexible and can easily be expanded. One can, for instance, imagine replacing the current isotropic spatial distribution component of the model to make it more intricate and parametrize the observed anisotropy of the system. It would also be straightforward to add more components to Equation~\ref{eq:model}, for instance to use the M31 dwarf galaxies to place constraints on the dark matter particle \citep[\eg][]{Kim2018} or the faint-end of galaxy formation \citet[\eg][]{Koposov2009}. Considering that, according to our inference, two thirds of dwarf galaxies are still undiscovered around M31, applying this framework to a wider and deeper M31 satellite sample (\eg thanks to the surveys that will be conducted with the Euclid, Rubin, or Roman telescopes), would yield important and tighter constraints on the global properties of this system and, possibly, on the dark matter properties. Finally, the method we describe could similarly be applied to other galaxies, either the MW itself \citep{Drlica2020_1}, or galaxies outside the Local Group \citep{Mao2021,Mutlu-Padkil2021,Carlsten2022,Nashimoto2022}. Reproducing the current analysis on the dwarf galaxy systems of other hosts will be essential to place robust constraints on cosmological and galaxy formation physics and to continue refining the use of dwarf galaxies as cosmological probes. This framework will be particularly useful considering the future arrival of the LSST and Euclid that are going to revolutionize the study of satellite systems beyond the Local Group.

\acknowledgements
We thank Denija Crnojevi\'c and Erik Tollerud for stimulating discussions about the work presented here.

Based on observations obtained with MegaPrime/ MegaCam, a joint project of CFHT and CEA/DAPNIA, at the Canada–France–Hawaii Telescope (CFHT), which is operated by the National Research Council (NRC) of Canada, the Institut National des Science de l’Univers of the Centre National de la Recherche Scientifique (CNRS) of France, and the University of Hawaii. 

The authors would like to acknowledge the High Performance Computing Center of the University of Strasbourg for supporting this work by providing scientific support and access to computing resources. Part of the computing resources were funded by the Equipex Equip@Meso project (Programme Investissements d'Avenir) and the CPER Alsacalcul/Big Data.

DW and AS were supported for this work by NASA through grants GO-13768, GO-15746, GO-15902, AR-16159, and GO-16273 from the Space Telescope Science Institute, which is operated by AURA, Inc., under NASA contract NAS5-26555.

GFT acknowledge support from the Agencia Estatal de Investigaci\'on (AEI) under grant {\it Ayudas a centros de excelencia Severo Ochoa convocatoria 2019} with reference CEX2019-000920-S, and from the Agencia Estatal de Investigaci\'on del Ministerio de Ciencia e Innovaci\'on (AEI-MCINN) under grant {\it En la frontera de la arqueolog\'ia galac\'actica: evoluc\'ion de la materia luminoso y obscura de la v\'ia L\'actea y las galaxias enenas del Grupo Local} with reference PID2020-118778GB-I00.

RI and NM acknowledge funding from the European Research Council (ERC) under the European Unions Horizon 2020 research and innovation programme (grant agreement No. 834148).

\bibliography{biblio.bib}
\bibliographystyle{apj}
\end{document}